\documentclass[english,gc, manuscript]{copernicus}
\usepackage[T1]{fontenc}
\usepackage{array}
\usepackage{longtable}
\usepackage{multirow}
\usepackage{graphicx}
\usepackage[authoryear]{natbib}

\makeatletter

\providecommand{\tabularnewline}{\\}

\usepackage{colortbl}
\definecolor{lightgray}{gray}{0.95}
\DeclareRobustCommand{\disambiguate}[3]{#2~#3}

\makeatother

\usepackage{babel}
\begin{document}
\title{``\textit{Thanks for helping me find my enthusiasm for physics!}''
The lasting impacts `research in schools' projects can have on students,
teachers, and schools}
\author[1,{*}]{Martin O. Archer}
\author[2,3]{Jennifer DeWitt}
\affil[1]{School of Physics and Astronomy, Queen Mary University of London,
London, UK}
\affil[2]{Institute of Education, University College London, London, UK}
\affil[3]{Independent Research and Evaluation Consultant, UK}
\affil[{*}]{now at: Space and Atmospheric Physics, Department of Physics, Imperial
College London, London, UK}
\correspondence{Martin O. Archer\\
(martin@martinarcher.co.uk)}
\runningtitle{Impact of PRiSE}
\runningauthor{Archer and DeWitt}
\maketitle
\nolinenumbers
\begin{abstract}
Using 6 years of evaluation data we assess the medium- and long-term
impacts upon a diverse range of students, teachers, and schools from
participating in a programme of protracted university-mentored projects
based in cutting-edge space science, astronomy, and particle physics
research. After having completed their 6-month-long projects, the
14--18 year-old school students report having substantially increased
in confidence relating to relevant scientific topics and methods as
well as having developed numerous skills, outcomes which are corroborated
by teachers. There is evidence that the projects helped increase students'
aspirations towards physics, whereas science aspirations (generally
high to begin with) were typically maintained or confirmed through
their involvement. Longitudinal evaluation 3~years later has revealed
that these projects have been lasting experiences for students which
they have benefited and drawn upon in their subsequent university
education. Data on students' destinations suggests that their involvement
in research projects has made them more likely to undertake physics
and STEM degrees than would otherwise be expected. Cases of co-created
novel physics research resulting from PRiSE also has seemed to have
a powerful effect, not only on the student co-authors but participating
students from other schools also. Teachers have also been positively
affected through participating, with the programme having influenced
their own knowledge, skills, and pedagogy, as well as having advantageous
effects felt across their wider schools. These impacts suggest that
similar `research in schools' initiatives may have a role to play
in aiding the increased uptake and diversity of physics and/or STEM
in higher education as well as meaningfully enhancing the STEM environment
within schools.
\end{abstract}

\introduction{}

Independent research projects provide extended opportunities for school
students to lead and tackle open-ended scientific investigations,
with so-called `research in schools' programmes, which have been emerging
in recent years, being a subset of these linked to current academic
(STEM) research \citep[e.g.][]{colle07,archer_report17,sousasilva17,iris20}.
It has been realised recently that, in general, more extended programmes
of STEM interventions with young people are required to be effective
in having lasting impact upon them compared to typical ``one off''
approaches (see the review of \citealp{archer20r4a} and references
therein). Independent research projects and `research in schools',
when appropriately developed and supported by expert mentors from
universities or industry, thus potentially align with this direction.

\citet{bennett16,bennett18} reviewed the evidence of impact from
independent research project endeavours globally. They found that
impacts on students were most often investigated, with various outcomes
being reported including improved understanding, practical and transferable
skills, or attitudes and aspirations towards science. \citet{dunlop19}
further suggest there is value in students participating in independent
research projects through developing their understanding about scientific
research/researchers and allowing them to make more informed decisions
about their future subject choices. It appears at present though that
only a few programmes consider the impact on students from under-represented
groups, presenting emerging evidence that increased engagement with
science can result from their involvement \citep{bennett16,bennett18}.
Finally, the review highlights that evaluations exploring the potential
long-term impacts of project work on students, such as subsequent
subject or career choices, are currently lacking.

While some studies into independent research projects use views from
teachers, these tend to little explore the impact on the teachers
themselves from their own participation in projects \citep{bennett16,bennett18}.
This aspect has recently been considered by \citet{rushton19} for
teachers engaged in projects with \citet{iris20}. Through in-depth
interviews with 17 research-active science teachers, they report that
through the projects they felt reconnected with science or research
and had developed as teachers, including in their pedagogy, skills
development, and recognition by school colleagues.

In general, however, \citet{bennett16,bennett18} note that with little
detail on the assessment criteria for independent research projects
being reported, further work is required to improve the quality of
evidence and to more fully explore the potential long-term benefits
of teachers' and students' (particularly those from under-represented
groups) involvement in independent research projects. This paper draws
on 6~years' worth of evaluation data from the `Physics Research in
School Environments' (PRiSE) programme of independent research projects
to explore the possible impacts on participating students, teachers
and schools.

\section{PRiSE}

`Physics Research in School Environments' (PRiSE) is a scalable framework
for independent research projects based in current physics research
that are mentored by active researchers \citep{archer_prise_framework20}.
The programme aims to equip 14--18 year-old school students (particularly
those from disadvantaged backgrounds) with the ability, confidence,
and skills in order to increase/sustain their aspirations towards
physics or more broadly STEM, ultimately enabling them to realise
these at higher education and thus contributing to increased uptake
and diversity of physics, and to some extent STEM \citep[cf.][]{archer20}.
Through working with teachers, PRiSE also aims to develop their professional
practice and build long-term university--school relationships that
raise the profile of science and mitigate biases/stereotypes associated
with physics within these schools, generally making them environments
which nurture and enhance all students' science capital \citep[cf.][]{iop14}. 

\begin{table*}
\begin{footnotesize}%
\begin{tabular}{>{\raggedright}p{0.25\columnwidth}lll>{\raggedright}p{0.33\columnwidth}}
\textbf{Project} & \textbf{Abbreviation} & \textbf{Years} & \textbf{Field} & \textbf{Description}\tabularnewline
\hline 
\noalign{\vskip6pt}
Scintillator Cosmic Ray Experiments into Atmospheric Muons & SCREAM & 2014--2020 & Cosmic Rays & Scintillator -- Photomuliplier Tube detector usage\tabularnewline
\noalign{\vskip6pt}
Magnetospheric Undulations Sonified Incorporating Citizen Scientists & MUSICS & 2015--2020 & Magnetospheric Physics & Listening to ultra-low frequency waves and analysing in audio software\tabularnewline
\noalign{\vskip6pt}
Planet Hunting with Python & PHwP & 2016--2020 & Exoplanetary Transits & Learning computer programming, applying this to NASA Kepler and TESS
data\tabularnewline
\noalign{\vskip6pt}
ATLAS Open Data & ATLAS & 2017--2020 & Particle Physics & Interacting through online tool with LHC statistical data on particle
collisions\tabularnewline
\end{tabular}\end{footnotesize}

\caption{A summary of the existing PRiSE projects at QMUL.\label{tab:projects}}

\end{table*}

PRiSE was developed at Queen Mary University of London (QMUL) in 2014,
where the four projects summarised in Table~\ref{tab:projects} currently
exist. These projects run for approximately six months from the start
of the UK academic year in September to just before the spring/Easter
break in March, culminating with students presenting their work at
a special conference held on university campus. Students typically
spend 1--2 hours per week throughout working on the project in groups,
usually outside of lesson time. Support from the university is provided
through workshops, school visits, monthly webinars, printed/multimedia
resources, and ad hoc emails as required. The provision within the
programme is explored in more detail in \citet{archer_prise_framework20}.
PRiSE has engaged a much more diverse set of school students and significantly
more disadvantaged groups than is typical. Furthermore it has been
found that students' success within the programme appears independent
of background, which has been attributed by teachers as due to the
extraordinary level of support offered \citep{archer_prise_schools20}.

The evaluation of PRiSE's pilot, which ran from 2014--2016 and involved
6~schools, suggested that students' awareness of current scientific
research, understanding of the scientific method, and skills were
enhanced by the programme and that teachers benefited through reconnecting
with their subject at an academic level, being challenged, and being
supported in their professional development \citep{archer_report17}.
The programme has grown significantly since then, having involved
67 London schools by 2020. This paper expands the evaluation of PRiSE's
potential impacts.

\section{Methods}

To evaluate the impact of the PRiSE programme, questionnaires were
distributed to students and their teachers at each year's student
conference held approximately 6~months after they started their PRiSE
projects. We were also able to contact a subset of individuals three
years after their participation. No control groups were established
due to ethical considerations, further explored in \citet{archer_prise_framework20},
which slightly limits this impact study. However, where possible we
draw from publicly available benchmark data.

\subsection{Instruments}

Paper questionnaires were handed out to students and teachers at our
student conferences each year (apart from 2020 where this was done
online due to the COVID-19 pandemic), which assessed the impact on
students and teachers at this 6-month stage. There were two questionnaires
for students. One of these was project-specific, relating to students'
confidence in scientific topics/practices relevant to their specific
project, i.e. those in Table~\ref{tab:projects}. The other applied
to PRiSE in general, asking students about skills development, aspirations,
and other ways they may have been affected by their involvement. Teachers
only completed a PRiSE-wide questionnaire, which not only asked for
their observations of impacts upon students but also how involvement
in the project has affected their own knowledge, skills, practice,
and wider school environments. The questions posed to both students
and teachers varied slightly from year-to-year and are found in Appendix~\ref{sec:6-month-questions}.
The PRiSE-wide questionnaires also included feedback on participants\textquoteright{}
experience of the programme, with this data forming the focus of a
separate paper \citep{archer_prise_framework20}.

These instruments were chosen in order to collect data from as wide
a range of students and teachers as possible as well as respecting
the limited time/resources of all involved (both on the school and
university sides). All data gathered was anonymous, with students
and teachers only indicating their school and which project they were
involved with. We have further anonymised the data by using pseudonyms
for the schools. More detailed information about the schools involved
can be found in \citet{archer_prise_schools20}. No protected characteristics
(such as gender or race) or sensitive information (such as socio-economic
background) were recorded. Our ethics statement on the forms informed
participants that the data was being collected for evaluation purposes
to determine the programme's impact, and that they could leave any
question they felt uncomfortable answering blank (also true of the
online form). 

For longer term evaluation, students were also asked on a separate
paper form at our conference to share their personal email addresses
so that we could follow up with them a few years later, in order to
explore potential lasting impacts of the programme. As with the main
questionnaires, this was presented as optional with an ethics statement
and description on how their data would be used clearly presented.
It was decided to contact cohorts of PRiSE students 3~years after
they started the project for this follow-up, so that students would
either be studying at university or at least (in the case of the youngest
PRiSE students) making university applications, hence giving us insight
into university destinations/plans. The students were emailed and
asked to fill out an online form, detailed in Appendix~\ref{sec:3-year-questions}.
The form contained primarily open-ended questions, to enable us to
understand PRiSE students\textquoteright{} long-term attitudes to
the programme, their higher education destinations, and what may have
affected these decisions. This sort of contextual data would not be
available by simply obtaining destination data from services such
as the Higher Education Access Tracker (HEAT, https://heat.ac.uk)
or requesting that schools provide it as a condition of their participation.
Doing so would have also risked schools declining to participate.
However, we do acknowledge that our approach reduces the number of
responses that could realistically be collected.

\subsection{Participants}

Data was collected from 153~students (aged 14--18) and 45~teachers
across 37 London schools. A breakdown of the number of responses per
year and how many schools these responses came from is given in Table~\ref{tab:response-rates}
where the total number of participants (and their schools) in attendance
at our conferences are also indicated. We note that, due to the COVID-19
pandemic, the 2019/20 programme was disrupted so we do not have reliable
information on how many students, teachers, and schools would have
successfully completed the programme that year. There is no indication
that the respondents differed in any substantive way from the wider
cohorts participating in the programme.

\begin{table}
\begin{centering}
\begin{tabular}{cc|cccccc}
 & \multicolumn{1}{c}{} & \textbf{2015} & \textbf{2016} & \textbf{2017} & \textbf{2018} & \textbf{2019} & \textbf{2020}\tabularnewline
\hline 
\multirow{2}{*}{\textbf{Students}} & Project-specific &  & 13/26 (50\%) & 30/70 (43\%) & 45/92 (49\%) & 40/97 (41\%) & \tabularnewline
 & PRiSE-wide &  & 13/26 (50\%) & 21/70 (30\%) & 46/92 (50\%) & 38/97 (39\%) & 35/?\tabularnewline
\hline 
\textbf{Teachers} & PRiSE-wide & 1/1 (100\%) & 6/6 (100\%) & 6/11 (55\%) & 9/16 (56\%) & 6/16 (38\%) & 17/?\tabularnewline
\hline 
\textbf{Schools} &  & 1/1 (100\%) & 6/6 (100\%) & 11/ 11 (100\%) & 13/15 (87\%) & 11/15 (73\%) & 19/?\tabularnewline
\hline 
\end{tabular}
\par\end{centering}
\caption{Response rates to questionnaires at PRiSE student conferences.\label{tab:response-rates}}

\end{table}

\subsection{Analysis}

Both qualitative and quantitative approaches were utilised in data
analysis, as the open and closed ended questions in the surveys generated
different types of data.

For all quantitative (numerical) data, uncertainties presented represent
standard (i.e. 68\%) confidence intervals. For proportions/probabilities
these are determined through the \citet{clopper34} method, a conservative
estimate based on the exact expression for the binomial distribution,
and therefore represent the expected variance due to counting statistics
only. Several statistical hypothesis tests are used with effect sizes
and two-tailed $p$-values being quoted, with a statistically significant
result being deemed as $p<0.05$. In general we opt to use nonparametric
tests as these are more conservative and suffer from fewer assumptions
(e.g. normality, interval-scaling) than their parametric equivalents
such as t-tests \citep{hollander99,gibbons11}. When comparing unpaired
samples a Wilcoxon rank-sum test is used, which tests whether one
sample is stochastically greater than the other (often interpreted
as a difference in medians). The Wilcoxon signed-rank test is used
to compare both a single sample to a hypothetical value or data from
paired samples to one another. Both versions test whether differences
in the data are symmetric about zero in rank. Finally, for proportions
we use a binomial test, an exact test based on the binomial distribution
of whether a sample proportion is different from a hypothesized value
\citep{howell07}. For ease of reference, further details about the
quantitative analyses are incorporated into the relevant sections
of the findings.

Thematic analysis \citep{Braun2006} was used to analyse the textual
(qualitative) responses. Instead of using pre-determined qualitative
codes to categorise the data, our analyses drew on a grounded theory
approach \citep{Robson2011,Silverman2010}, letting the themes emerge
from the data itself. This process involved the following steps:
\begin{enumerate}
\item Familiarisation: Responses are read and initial thoughts noted.
\item Induction: Initial codes are generated based on review of the data.
\item Thematic Review: Codes are used to generate broad themes (which we
refer to as dimensions) and identify associated data.
\item Application: Codes are reviewed through application to the full data
set.
\item Reliability: Codes are applied to a subset of data by second coder
to check reliability.
\item Final Coding: Final codes are applied to the data. 
\item Analysis: Thematic overview of the data is confirmed, with examples
chosen from the data to illustrate the themes (dimensions).
\end{enumerate}
Overall there was $92\pm2\%$ agreement between the two coders, which
corresponds to a Cohen\textquoteright s kappa of 0.836 (Cohen\textquoteright s
kappa is unity minus the ratio of observed disagreement to that expected
by chance, hence ranges from 0 to 1, \citealp[e.g.][]{mchugh12}).
Disagreements were resolved by discussion to arrive at the final coding
presented in the paper. Our analyses examine the impact on students
and teachers from schools which completed the programme, as it was
not possible to gather evidence from schools which dropped out during
the year. Future work could attempt to investigate this, subject to
funding and ethical approval.

\section{Impact on students}

This first section of the findings examines the impact of PRiSE on
students at the 6-month (captured at our student conferences) and
3-year (captured online) stages. Impact on students through the co-production
of research is also briefly discussed.

\subsection{6-month stage evaluation}

We assess the impact on students in three broad areas related to the
aims of the programme: their confidence in scientific topics and methods;
their skillsets; and their aspirations towards pursuing physics or
STEM.

\subsubsection{Confidence}

At our conferences from 2016--2019 students ($n=127$) were asked
to rate their confidence (using a 6-point Likert scale) in topics
and practices relevant to their projects (see Table~\ref{tab:topics-practises}).
Additionally they were asked to retrospectively assess their confidence
before having undertaken the project. Given that the options varied
by project, this was performed on separate paper questionnaires to
the more general PRiSE-wide questionnaire that applied to all projects,
however, the same level of anonymity and ethical considerations were
afforded to students here too. While for a subset of students \citet{archer_report17}
also surveyed students before undertaking projects, we opted not to
continue this. This is firstly because surveys taken on different
days necessarily include natural intraindividual variability \citep{eid99}
which cannot be separated out from any real stable change with only
two survey points. Secondly, a retrospective assessment following
the project also means that an apparent decrease in confidence, known
as the \citet{kruger99} effect, is avoided because prior to the interventions
confidence may have been artificially high as individuals are not
aware of what they do not know. Given that many other repeated STEM
intervention programmes have resulted in no overall changes from before
to after \citep[e.g.][]{shattering17,connect18}, we set the benchmark
to be a statistically significant positive change.

\begin{table}
\begin{centering}
\begin{tabular}{ccccc}
\multicolumn{4}{c}{\textbf{Topics}} & \textbf{Practices}\tabularnewline
SCREAM & MUSICS & PHwP & ATLAS & \tabularnewline
\hline 
Neutrinos & Plasma & Planets & Fundamental Particles & Mathematical Models\tabularnewline
Muons & Magnetic Fields & Stars & Fundamental Forces & Experiment Design\tabularnewline
Cosmic Rays & Space & Gravity & Particle Detectors & Calibration (SCREAM)\tabularnewline
Particle Detectors & Magnetosphere & Exoplanet Detection & Particle Interactions & Statistical Analysis\tabularnewline
Anti-particles & Waves &  &  & Error Analysis\tabularnewline
Special Relativity & Resonance &  &  & Drawing Conclusions\tabularnewline
 &  &  &  & Collaborating\tabularnewline
 &  &  &  & Presenting\tabularnewline
 &  &  &  & Writing\tabularnewline
 &  &  &  & Reviewing Literature\tabularnewline
 &  &  &  & Programming (PHwP)\tabularnewline
\end{tabular}
\par\end{centering}
\caption{The scientific topics and practices used in assessing students' confidence.\label{tab:topics-practises}}
\end{table}

\begin{figure*}
\begin{centering}
\includegraphics{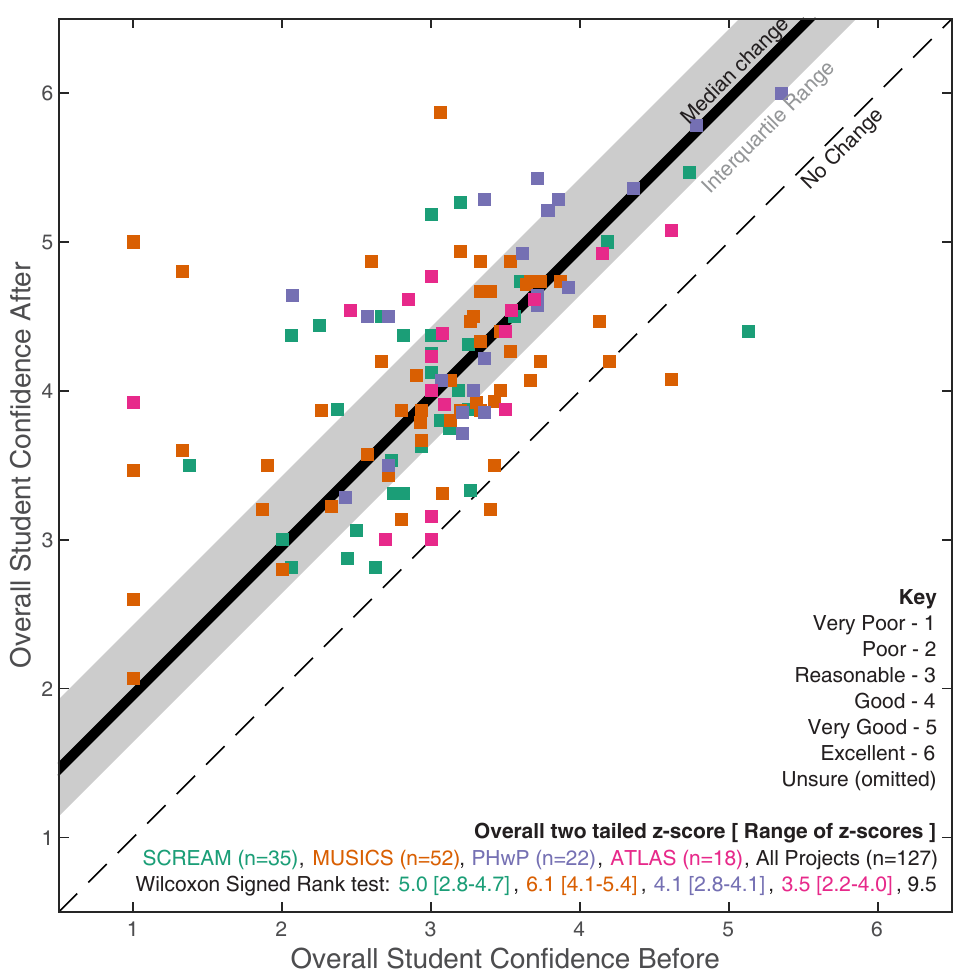}
\par\end{centering}
\caption{Overall student confidence in relevant scientific topics/methods before
and after taking part in PRiSE ($n=127$). Data points are coloured
by project. Overall changes are also indicated through the interquartile
range (grey area) and standard confidence interval in median (black
bar). Overall and ranges of $z$-scores in two-tailed Wilcoxon signed-rank
tests are also listed by project.\label{fig:confidence}}
\end{figure*}

We test the paired before and after data for each topic and practice,
omitting any where students listed either as unsure, using a Wilcoxon
signed-rank test. The results show statistically significant increases
for all the topics and practices, with the range of two-tailed $z$-scores
by project listed in square brackets in Figure~\ref{fig:confidence}
($z\geq1.96$ corresponds to 95\% confidence or more). To give an
overall measure of these changes, we take an average for each student
across all topics and practices (again omitting unsures as before)
and these are plotted in Figure~\ref{fig:confidence} showing $96\pm2\%$
of students have increased in confidence. The median overall change
across all projects was $0.92\pm0.04$ points, indicated as the black
bar in Figure~\ref{fig:confidence} along with interquartile range
(grey area), whereas the mean was slightly higher at $1.08\pm0.06$
due to a positive skewness (the uncertainty refers to the standard
error in the mean). The overall results show positive changes across
all projects to a high level of confidence, as indicated in the figure,
with no real variation in results between the different projects or
between schools. Therefore, students' confidence in scientific topics
and methods seems to have substantially increased as a result of PRiSE
and almost all students reported this benefit. This gain in confidence
has also been noted in teachers' comments:\begin{quote} ``\textit{They
have become more confident in communicating their ideas and realised
that they are not too young to do research.}'' (Teacher~1, Hogwarts,
SCREAM 2015)\\
\textit{``This has been a challenging experience for the students
taking part. Students have gained a better appreciation of real science
and built confidence}.'' (Teacher~3, Xavier's Institute for Higher
Learning, MUSICS 2016)\end{quote}

\subsubsection{Skills}

\begin{figure}
\begin{centering}
\includegraphics{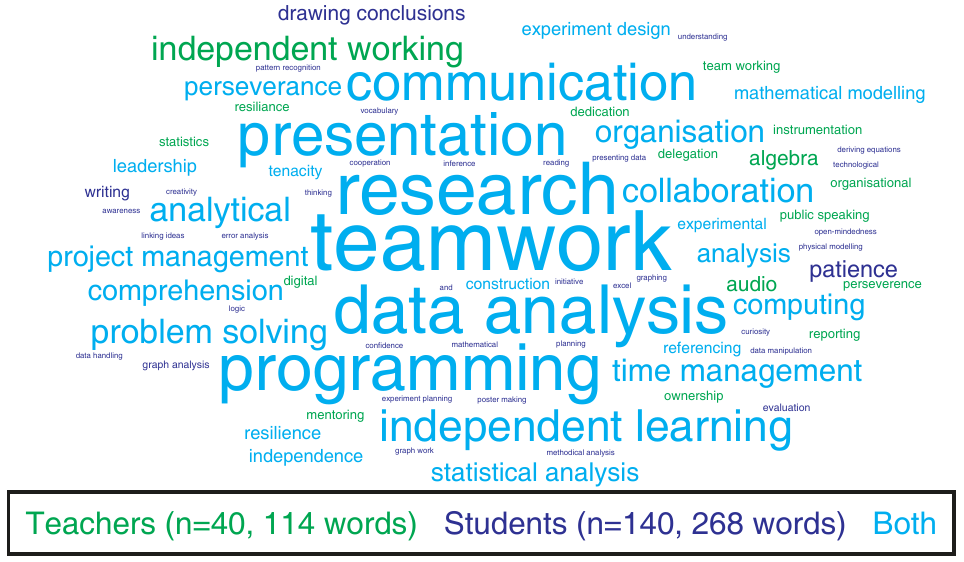}
\par\end{centering}
\caption{Word cloud of skills developed by students. Colours indicate words
identified by students (blue), teachers (green), or both (cyan). Students
and teachers have been given equal total weight.\label{fig:keywords}}

\end{figure}

From 2016 onwards we asked students ($n=140$) to list which skills,
if any, they felt they had developed through their PRiSE project.
Teachers ($n=40$) were also asked to indicate their observations
on skills that their students had developed during the programme.
We extracted keywords from any prose responses and sifted through
the data performing keyword clustering. This latter step involved
identifying synonymous skills and relabelling them so there was a
consistency of terminology throughout. This processing resulted in
a dataset of 79 unique skills. Students tended to identify on average
around 2 distinct skills each whereas teachers typically listed 3,
though the responses per person ranged up to 5 and 6 respectively.
Figure~\ref{fig:keywords} shows the skills identified as a word
cloud, where students and teachers have been given equal weight by
normalising their counts by their respective totals. Colours indicate
from whom the words originated, showing a large amount of agreement
between the two groups. All the skills listed are highly relevant
to being a scientist, with the most cited being (in descending order)
teamwork, research, data analysis, programming and presentation skills.
These results remain fairly consistent with those from the pilot and
arguably indicate areas where university / research physics differ
substantially from the regular school experience. Skills development
was not mentioned in any of the PRiSE projects' resources (e.g. there
was no list of ``skills to be developed'') so we are confident in
the validity of these results, though we cannot rule out that teachers
may have influenced students' answers. Therefore, through experiencing
and being involved in research-level physics, it seems likely students
have gained new, or further developed existing, skills, constituting
a positive impact upon them. This has been further expanded upon in
the teacher feedback:\begin{quote}``\textit{They have developed
presenting skills, they do get that }{[}at school{]}\textit{ but not
for academic poster sessions. The unique skills from the project were
the exposure to the physics, analysis, independence; it has allowed
them to access the world.}'' (Teacher~1, Hogwarts, SCREAM 2015)\\
``\textit{Challenging opportunity }{[}for students{]}\textit{ to
broaden skills and experience.}'' (Teacher~23, Smeltings, ATLAS
2019)\\
``\textit{Great for developing pupil research skills and getting
confident in cross referencing scientific articles, a very important
skill for them in post-college education.}'' (Teacher~42, Hill Valley
High School, ATLAS 2020)\end{quote}

\subsubsection{Aspirations\label{subsec:Aspirations}}

To assess whether students' aspirations were affected at the 6-month
stage, we first undertook a qualitative analysis in 2018--2019 asking
in an open question how students' thoughts about future subject choices
or careers might have been affected through doing the project. A thematic
analysis of the 63 responses revealed three distinct dimensions, each
with their own set of underlying codes. The first of these concerned
how much the students' felt that they had wanted to study either physics
or a STEM subject before even undertaking the project ($n=22$), which
tended to be raised if they already wanted to pursue this route or
if they were not interested in these subjects. The second dimension
covers the students' aspirations following the project ($n=13$),
revealing that students were either now wanting to pursue physics/STEM,
were considering these as potential options, or were simply unsure.
Finally, the third dimension was an expression of change as a result
of their involvement in PRiSE ($n=22$), which typically stated that
it had confirmed their subject choice, made them more likely to pursue
physics/STEM, had not affected them, or (in a small number of cases)
had deterred them from continuing with physics/STEM. These themes
align with the findings of \citet{dunlop19} on how independent research
projects can affect students' aspirations.

We show the dimensions and codes in Table~\ref{tab:aspirations-coding},
also giving counts of the number of responses which fall within them
\citep[cf.][]{sandelowski01,sandelowski09,maxwell10}. We note that
some students' responses covered more than one of the dimensions,
but none spanned all three. Out of the 63 responses to this question,
11 did not fit into any of these three dimensions, instead highlighting
aspects of the programme they enjoyed (research, teamwork, real applications
of physics, and what physics at university is like) but not explicitly
stating their subject aspirations or how they may have been changed
by the project. From these counts, it is clear that in both dimensions
2 and 3 the totals indicating positive effects from PRiSE are greater
than the neutral or negative responses. However, given that these
numbers are rather small and derived from a qualitative coding we
do not attempt to make a statistical interpretation about the impact
of PRiSE on students' physics/STEM aspirations based on them.

\begin{table*}
\begin{footnotesize}%
\begin{tabular}{c|>{\centering}p{0.12\columnwidth}>{\centering}p{0.12\columnwidth}>{\centering}p{0.12\columnwidth}>{\centering}p{0.12\columnwidth}}
\multicolumn{5}{l}{\textbf{Dimension 1: Likeliness of wanting to study physics/STEM before
the project}}\tabularnewline
\hline 
\textbf{Codes} & Wanted to already & Unrelated to future choices &  & \tabularnewline
\textbf{Count} & 16 & 6 &  & \tabularnewline
\multicolumn{1}{c}{} &  &  &  & \tabularnewline
\multicolumn{5}{l}{\textbf{Dimension 2: Likeliness of wanting to study physics/STEM after
doing project}}\tabularnewline
\hline 
\textbf{Codes} & Now wants to & Now considering as option & Unsure on future choices & \tabularnewline
\textbf{Count} & 4 & 3 & 6 & \tabularnewline
\multicolumn{1}{c}{} &  &  &  & \tabularnewline
\multicolumn{5}{l}{\textbf{Dimension 3: Change in likeliness of wanting to study physics/STEM
due to project}}\tabularnewline
\hline 
\textbf{Codes} & Confirmed as subject choice & More likely & Didn't change mind & Deterred\tabularnewline
\textbf{Count} & 6 & 6 & 8 & 2\tabularnewline
\end{tabular}\end{footnotesize}

\caption{Qualitative coding of students' responses concerning their future
aspirations along with counts ($n=52$ of $63$).\label{tab:aspirations-coding}}

\end{table*}

\begin{figure*}
\begin{centering}
\includegraphics{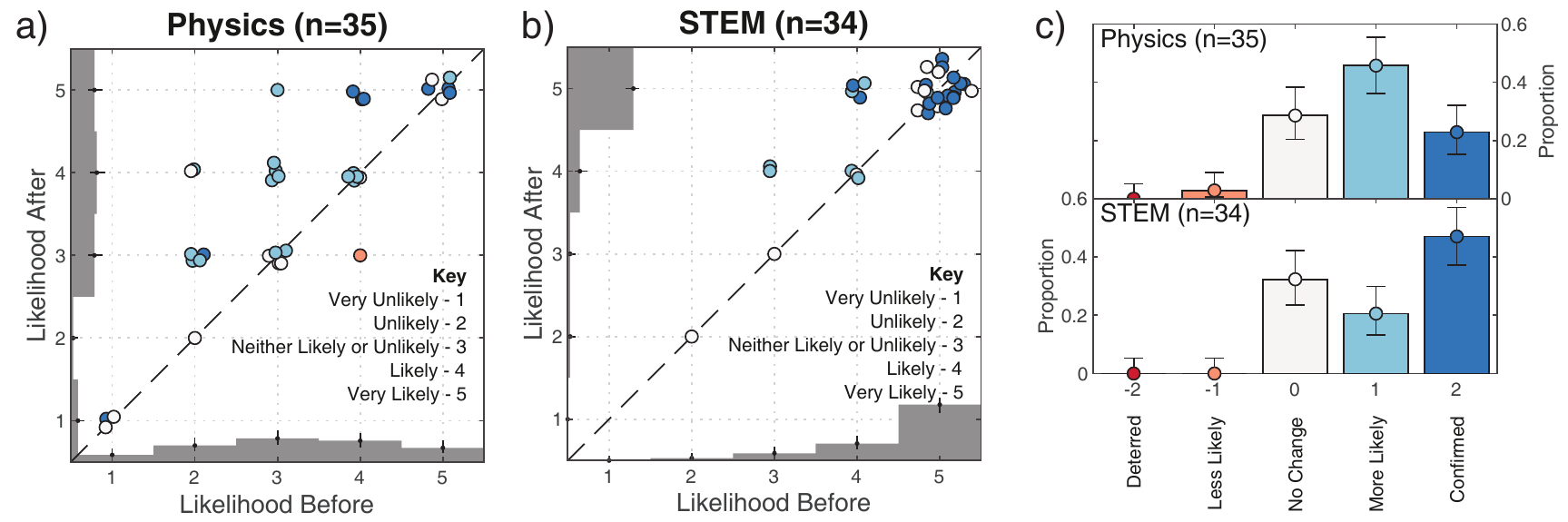}
\par\end{centering}
\caption{Likelihood before and after of PRiSE students continuing with a) physics
and b) STEM education. Datapoints have been jittered for visibility.
Marginal distributions (grey areas) are also shown for both before
and after. How PRiSE specifically has affected students' aspirations
are shown in panel c for physics (top) and STEM (bottom), where the
colours used are also reflected in panels a and b. Error bars denote
the standard ($1\sigma$) \citet{clopper34} interval.\label{fig:aspirations}}
\end{figure*}

Instead, informed by these promising preliminary results, we implemented
in 2020 a quantitative approach to assessing how PRiSE may have affected
students' aspirations. In a similar manner and with similar justification
to our evaluation of students' confidence, we asked students ($n=35$)
to assess their likelihood (using a 5-point Likert scale) of continuing
with physics and STEM as well as retrospectively assessing these from
before the project. We shall call this the `absolute scale', since
it pertains to students' absolute likelihood of continuing these subjects,
and code it to values of 1--5 as shown in Figure~\ref{fig:aspirations}a--b.
In addition, we asked how working on the project affected their thoughts
on physics and STEM as future subject choices using a different 5-point
scale, where the wording of the options used were informed by the
previous qualitative results in Table~\ref{tab:aspirations-coding}.
We refer to this as the `relative scale', since it concerns whether
the students feel their thoughts changed as a result of PRiSE, and
code it to values from -2 to +2, as shown in Figure~\ref{fig:aspirations}c.
One student only answered these questions relating to physics but
not for STEM.

When considering physics aspirations, there was no clear positive
bias towards the subject beforehand (see horizontal distribution in
Figure~\ref{fig:aspirations}a) with a mean value on the absolute
scale of $3.23\pm0.21$ ($p=0.264$ in a one-sample Wilcoxon signed-rank
test against null hypothesis of 3). The vertical distribution shows
some shift towards greater values on the absolute scale after the
projects, now exhibiting overall positive results (mean of $3.69\pm0.20$,
$p=0.006$). From the paired data, $40\pm10\%$ of students increased
in likelihood of studying physics on the absolute scale (though no
students who were very unlikely before showed any increase) and only
one student's likelihood decreased (to a neutral stance), with the
mean change being $+0.46\pm0.12$. While this indicates only moderate
changes in students' absolute physics aspirations, they are statistically
significant ($p=8\times10^{-4}$ in a Wilcoxon signed-rank test).
On the relative scale displayed in the top panel of Figure~\ref{fig:aspirations}c,
however, $69\pm9\%$ of students report that the projects either made
them more likely to continue with physics or confirmed it, with again
only one student becaming less likely to pursue physics. The average
was $+0.89\pm0.13$, greater than zero with high confidence ($p=2\times10^{-5}$).
No trends were present by project or school. Since we colour the datapoints
in panel~a of Figure~\ref{fig:aspirations} based on the students'
responses on the relative scale (panel~c), it is clear that most
of the students whose absolute physics aspirations increased attribute
this in some way to PRiSE, while around half of those which did not
increase on the absolute scale still claim positive influence by PRiSE.

Students' STEM aspirations, unlike physics, were already incredibly
high before the projects as shown in Figure~\ref{fig:aspirations}b
with a mean value on the absolute scale of $4.53\pm0.14$. Because
of this, a large proportion ($67\pm9\%$) of students would be unable
to increase in value on the absolute scale due to already giving the
highest rating. No students decreased in likelihood on the absolute
scale and only 6~students out of the 34 increased, though this constituted
around half of the students who could possibly increase (gave a 4
or below beforehand). This slight change in the paired data (average
$+0.18\pm0.07$ across all students) was still statistically significant
($p=0.031$). The bottom panel of Figure~\ref{fig:aspirations}c
reveals a bimodal distribution on the relative scale of whether PRiSE
affected students' STEM aspirations, with the largest peak at +2 (it
confirmed their subject choice) and a smaller one at 0 (no change).
No students reported being less likely to pursue STEM due to PRiSE.
$68\pm9\%$ of students indicated PRiSE's likely positive influence
on their STEM aspirations (a result similar to physics aspirations),
with the mean being $1.15\pm0.15$ which is again a clear positive
result ($p=1\times10^{-5}$). As before there was no real variation
in these results by project or school. All students who increased
on the absolute scale attribute this to PRiSE, whereas a majority
who did not change still indicate PRiSE had a positive effect on their
STEM aspirations.

Ideally one would benchmark the likelihoods before PRiSE against larger
surveys of similarly-aged students' aspirations to test whether PRiSE
students were more likely to continue with physics/STEM anyway. Unfortunately,
direct comparisons are not possible due to the differing ways the
relevant questions have been structured across national surveys. However,
such research has shown that STEM degree aspirations amongst all students
remains similar to the makeup of STEM vs. non-STEM A-Level subject
choices, implying that almost all students studying at least one STEM
A-Level likely aspire towards a STEM degree \citep{wellcome19}. Furthermore,
science aspirations are highly correlated to `science capital' and
begin to form at an early age \citep{moote20}. Therefore, it is not
surprising that PRiSE students' likelihood of wanting to continue
with STEM was high beforehand.

The follow up qualitative question, asking students to explain how
or why their thoughts about subject choices had been affected by the
project typically mentioned how their interest, enjoyment, or understanding
had been enhanced, e.g.\begin{quote} ``\textit{Before working on
Planet Hunting With Python, I was already quite focused on studying
in a STEM field, and the main reason I signed up for the project was
because of my interest in physics. After the project, I felt as though
my decision to pursue such an area was only further cemented.}''
(Student~120, Octavian Country Day School, PHwP 2020)\end{quote}The
one student who reported being less likely to pursue physics (but
was not affected with regards to STEM) noted that\begin{quote} ``\textit{I
already had my mind set on doing STEM subjects at university, but
now I am less interested in physics as I have come to see how some
areas are more challenging than others and I wouldn't want to specialise
in those areas.}'' (Student~136, Jedi Academy, ATLAS 2020)\end{quote}In
light of this response, arguably the project enabled this student
to make a more informed decision \citep[cf.][]{dunlop19} rather than
necessarily causing harm. Some students raised the idea of their science
identity affecting their likelihood beforehand, a known key factor
in students' aspirations \citep[e.g.][]{aspires13,aspires20}, with
PRiSE showing mixed results in affecting this\begin{quote}``\textit{I
never really saw Physics as a choice for me, as I did not want to
do it, and the project hasn't changed my mind about this.}'' (Student~133,
Imperial Academy, ATLAS 2020)\\
``\textit{I am not very great at physics but this project made me
more interested and invested.}'' (Student 135, Xavier's Institute
for Higher Learning, SCREAM 2020)\end{quote}Another theme that emerged
in response to this question was that PRiSE added something not usually
accessible to them in school, the research, which helped cement students'
subject choices and potentially influenced their career aspirations
\citep[cf.][]{archer20}\begin{quote}``\textit{Because I've always
wanted to go into research and this project showed me that I enjoy
it.}'' (Student~138, Jedi Academy, ATLAS 2020)\\
``\textit{I've always assumed that I would work towards studying
physics since GCSE; this project was very useful in seeing what research
may be like if I took that path after education.}'' (Student~143,
Sunnydale High School, MUSICS 2020)\\
``\textit{I was always interested in maths and physics, especially
maths. However this project showed me what we do not do in physics
lessons, the research. For me this is one of the most important things
in physics.}'' (Student~145, Sunnydale High School, MUSICS 2020)\end{quote}

Students' aspirations have been found to be extremely resilient and
very difficult to change, with most (even protracted series of) interventions
yielding no statistically significant overall effects \citep{aspires13,archer14,archer20r4a}.
With this in mind, the moderate increase in physics and the slight
positive change in STEM absolute aspirations are considerable compared
to the sector and what is realistically achievable with the resources
afforded. The relative scale indicates that students feel that PRiSE
has influenced their subject choices, typically either confirming
or making them more likely to follow either physics or STEM, with
$80\pm8\%$ of students indicating a positive effect in either or
both.

\subsection{3-year stage evaluation}

To date we have undertaken long-term evaluation for three cohorts
of PRiSE students who had participated in the academic years 2015/16
(cohort~1), 2016/17 (cohort~2), and 2017/18 (cohort~3). At our
student conferences 72~students from these three cohorts left contact
details with us for this purpose, which were well spread across the
different schools involved. Across the three cohorts, the bounce rate
was $46\pm7\%$ (predominantly due to now inactive school email addresses
being given) and the emails were opened by 24 PRiSE students (a rate
of $62\pm9\%$ from non-bouncing emails, again well spread across
different schools) with 14 filling out the survey though we purposely
cannot identify individuals from responses. While this is a relatively
small number of responses, longitudinal evaluation is notoriously
difficult for under-resourced engagement programmes \citep[e.g.][]{archer20r4a}
and there are still significant and useful results from the data,
which we present in this section. The evaluation covers the perceived
legacy of PRiSE on these students, as well as the students' higher
education destinations and the factors which may have affected these.

\subsubsection{Legacy}

Most of the 14 PRiSE students who responded were aged 16--17 when
they undertook their projects, with two students aged 15--16, and
one student each in the ranges 14--15 and 17--18. All of them remembered
undertaking a physics research project with the university. When asked
what experiences they remembered from the project, the 11 open responses
provided (3 students did not answer this or the following question)
could be categorised as concerning the underlying science\begin{quote}
``{[}It{]}\textit{ helped us to really solidify our understanding
of harmonics.}'' (Student~2, cohort~1)\\
``\textit{Learning about how the magnetosphere works and its importance.}''
(Student~8, cohort~2)\\
``{[}I{]}\textit{ learned about exoplanets.}'' (Student~14, cohort~3)\end{quote}the
process of undertaking university/research style work\begin{quote}
``\textit{It was very intriguing to have played with actual data
and have an attempt at analysing the sound wave forms that we were
given.}'' (Student~2, cohort~1)\\
``\textit{Setting up the experiments and working through problems
as they arose.}'' (Student~7, cohort~2)\\
``\textit{Working with other students on a project we did not know
much about before and presenting it in front of a lot of people.}''
(Student~11, cohort~3)\\
``\textit{Creating a formula to use in order to calculate {[}the{]}
surface area of a scintillator which is hit directly by muons, {[}and{]}
observing building schematics to find how much matter muons pass through
during travel into }{[}the{]}\textit{ building.}'' (Student~12,
cohort~3)\\
``\textit{I remember getting to experience some more advanced practical
physics that was more reminiscent of university lab work than school
lab work.}'' (Student~13, cohort~3)\end{quote}the skills they
developed through their involvement\begin{quote} ``{[}I{]}\textit{
learned how to analyse sound in audacity.}'' (Student~5, cohort~2)\\
``\textit{I built good teamworking and project management skills}''
(Student~11, cohort~3)\\
``\textit{Learning to code in Python.}'' (Student~14, cohort~3)\end{quote}as
well as having enjoyed the overall experience\begin{quote} ``\textit{having
a great time and meeting some lovely people}'' (Student~13, cohort~3)\end{quote}When
asked how they have used the experiences since, if at all, of the
same 11 respondents all bar two provided examples. These concerned
their skills development\begin{quote} ``{[}It{]}\textit{ really
helped to develop our teamwork skills, which I have used frequently
in most things that I do in my academic education. Also there is a
huge element problem-solving and how to undertake the project/study
is fundamental in my Engineering degree for electronics, I have used
it a lot.}'' (Student~2, cohort~1)\\
``\textit{Now I }{[}have{]}\textit{ used python in my computational
physics module at university.}'' (Student~3, cohort~1)\\
``\textit{I am currently at university doing many group projects.
Taking part in the Queen Mary magnetosphere project has helped me
improve my team building and communication skills.}'' (Student~8,
cohort~2)\\
``\textit{The presentation helped me improve my public speaking and
speaking confidence.}'' (Student~13, cohort~3)\end{quote}as well
as activities they have done since\begin{quote} ``\textit{It was
a good introduction to conducting experiments. I carried out a CREST
gold research project after this experience.}'' (Student~7, cohort~2)\\
``\textit{I've tried to be confident to defend a project in front
of people and to show an inquisitive attitude.}'' (Student~11, cohort~3)\\
``\textit{I now study architecture, so the observing building schematics
and 3D mathematics were both useful experiences.}'' (Student~12,
cohort~3)\\
``\textit{The lab work was useful as it gave me an idea of how to
work in a uni lab, which is particularly helpful now that I am at
uni.}'' (Student~13, cohort~3)\end{quote}Of the two negative responses,
both stating ``\textit{I haven't}'', one caveated though that ``\textit{but
that's just because I haven't had to do a group project since}''
(Student~9, cohort~3). Overall, students' responses suggest that
their PRiSE projects were lasting and beneficial experiences that
they have been able to draw from in their subsequent educational activities
and development.

\subsubsection{Destinations\label{subsec:Destinations}}

\begin{figure}
\begin{centering}
\includegraphics{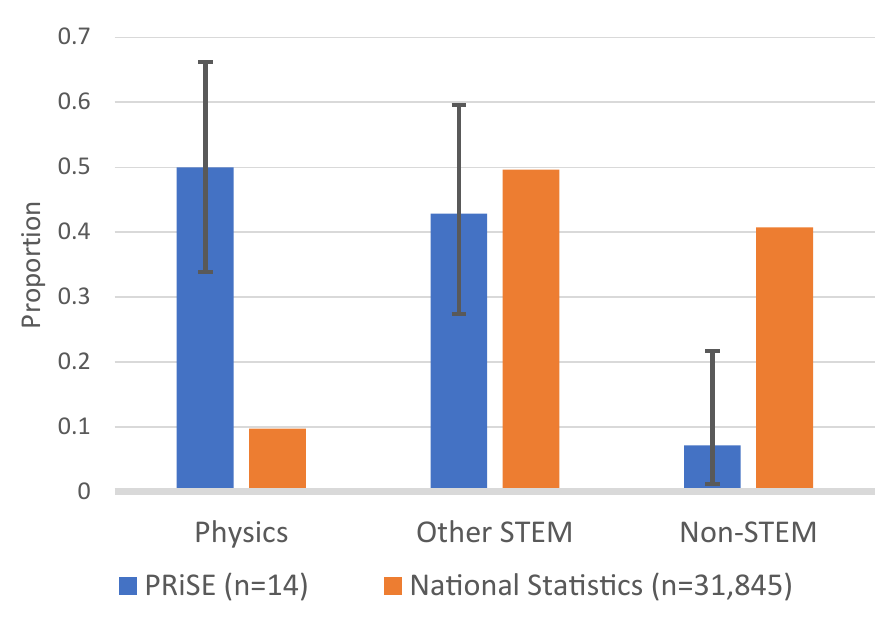}
\par\end{centering}
\caption{Degree destinations of PRiSE students (blue) compared to UK national
statistics of A-Level physics students (orange). Error bars denote
the standard (1$\sigma$) \citet{clopper34} intervals.\label{fig:destinations}}
\end{figure}

All the PRiSE students reported that they were studying at university
when the survey was conducted, apart from one who due to their age
(they were 14--15 when involved in PRiSE) intended to. We asked the
students what subject they were studying at university (or planned
to study in the case of the one student) giving the options of physics
(or combination including physics), another STEM subject, or a non-STEM
degree. The results of this are shown in blue in Figure~\ref{fig:destinations},
where we compare these to the degree destinations of physics A-Level
students nationally \citep{iop12} in orange (this data only includes
students that went on to higher education). The 7 out of 14 PRiSE
students going on to study a physics degree is considerably higher
than the national rate of $9.7\%$ ($p=1\times10^{-4}$ in a binomial
test). While the number of students going on to study other STEM subjects
(6 out of 14) are consistent with the national statistics, the increased
uptake of physics leads to the overall STEM-degree proportion (13
out of 14) also being significantly greater than found nationally
($59.3\%$, $p=0.012$). Given the diversity of schools involved (discussed
in-depth in \citealp{archer_prise_schools20}) and the known barriers
to STEM higher education for underrepresented groups \citep[e.g.][]{case14,wellcome19},
it is highly unlikely these results can be explained simply by PRiSE
schools tending to produce more physics and STEM students anyway.
Another consideration may be that PRiSE students were already highly
likely to continue their physics education beforehand anyway. While
this did not appear to be the case for the 2019/20 cohort discussed
in section~\ref{subsec:Aspirations}, with it being shown that PRiSE
led to some increased physics and STEM aspirations, the destinations
results here came from different cohorts so this remains a possibility.
However, given that PRiSE has been a consistent framework (both in
terms of schools targeting and delivery) throughout, it is reasonable
to assume that the 2019/20 cohort is representative of the others
used here and thus may be used as comparable samples. This then suggests
that students' involvement in PRiSE may very well make them more likely
to pursue physics and STEM degrees.

The students' reasons behind choosing their degree subjects and what
influenced them varied. We note that one student out of the 12 that
responded to these questions referenced the research project (which
was not prompted in the question)\begin{quote}``\textit{I did the
sounds of space project that you organised a couple of years ago and
am now pursuing a physics degree from Cambridge. Thanks for helping
me find my enthusiasm for physics!}'' (Student~4, cohort~1)\end{quote}However,
in general the responses were quite brief and did not give much insight
into the likely many factors which may have played a role in their
subject choices \citep[cf.][]{aspires13,aspires20}. While for cohorts
2 and 3 we added questions that explicitly asked about PRiSE's influence
on the students' degree choices, the small number of responses and
brief answers simply highlight the need for more in-depth qualitative
longitudinal evaluation, though this is challenging to undertake.
We note that there is little research into engagement programmes'
effectiveness at influencing students' destinations, both correlative
and causal, as raised in a recent review \citep{robinson20}. However,
given that PRiSE shows some statistically significant positive effects
on students' aspirations, which are known to be highly resilient,
at the 6-month stage as well as PRiSE students being more likely to
study physics and STEM degrees than would otherwise be expected (further
statistically significant effects), the results of this empirical
enquiry into PRiSE's potential effect on students' destinations are
perhaps promising.

\subsection{Co-production of research}

Co-producing publishable physics research between researchers and
school students is not an explicit aim of PRiSE, unlike for instance
the more researcher-driven ORBYTS programme \citep{sousasilva17}.
The rationale behind this is explored further in \citet{archer_prise_framework20}.
Nonetheless, in a few instances genuinely novel preliminary results
have come from students' independently motivated work on the MUSICS
project (the other PRiSE projects are unlikely to result in publishable
physics research). The first of these originated from the 2016/17
cohort, where a group from a girls' school in an area of particularly
high deprivation discovered a series of decreasing-pitch ``whistle''
sounds which lasted several seconds (corresponding to 5 days in reality).
Through collaboration with the students and further investigation
by professional scientists, it was discovered that these unexpected
sounds corresponded to the natural oscillations of Earth's magnetic
field lines following a solar storm. Such wave events had been deemed
rare previously but, due to the accessibility of exploring the sonified
data, were found to in fact be quite common thanks to the students'
discovery. The work was presented at several international scientific
conferences and eventually published in the journal `Space Weather'
with the students and their teacher listed as co-authors \citep{archer18}.
Information about these developments resulting from the students'
project work was continually passed on via their teacher, who in turn
responded with the students' comments:\begin{quote}``\textit{It
was a very rewarding experience which allowed us an insight into the
research conducted at university level. This helped us to develop
crucial skills needed in the next years of our studies. It was truly
amazing to hear how significant the event we found was and that it
will be forming the basis of a proper scientific paper.}''\\
``\textit{Being a part of the university's research and the subsequent
paper published is truly an amazing opportunity. It was really interesting
to find such a significant event and we gained so much experience
and developed many skills during our research that will be useful
in our university careers.}''\end{quote}The publication garnered
widespread media attention, for example featuring on BBC Radio 4's
`Inside Science'. Unfortunately we were unable to find out how the
news of the publication had been shared across the school involved
and affected other students' thoughts about physics. However, through
publicising the result across all schools involved in PRiSE via teachers,
it appears to have had a powerful effect on PRiSE students at other
schools:\begin{quote}\textquotedblleft \textit{Hearing that other
kids at other schools have actually produced a paper, it just gives
you hope that it\textquoteright s actually something I can do.}\textquotedblright{}
(Student, mixed state school in area of particularly high deprivation,
MUSICS, BBC Radio 4 interview, Oct 2018),\end{quote}potentially highlighting
through demonstration by their peers that (research-level) physics
is something which is accessible to `people like me', thereby breaking
down known barriers to participation in physics and science generally.

While no other publications have resulted as of yet, a group of students
(from a selective boys' academy) in 2018/19 identified undocumented
instrumental noise present in the data which researchers and satellite
operators were unaware of. Another group (from a non-selective mixed
academy with particularly high free school meal percentages and in
an area of particularly high deprivation) in 2019/20 decided to investigate
the relationship between the recently discovered aurora-like STEVE
(Strong Thermal Emission Velocity Enhancement) phenomenon \citep{macdonald18}
and magnetospheric ULF waves, finding some differences in wave activity
during published STEVE events to typical levels, which might lead
to promising results with significant additional work. Finally two
more groups in 2019/20 (a non-selective mixed academy; and a partnership
between an independent girls' school and three mixed state schools,
one with particularly high free school meal percentages and in an
area of particularly high deprivation ) working on a follow-up campaign
to the previous students' co-authored paper uncovered other wave events
during solar storms with novel features that are currently being investigated
further by professional scientists and may be publishable in the future.

\section{Impact on teachers and schools}

Possible impacts upon teachers and schools from PRiSE were first explored
using qualitative responses to open-ended questions and then further
investigated with quantitative data gathered in 2019-2020.

\subsection{Thematic analysis}

From a thematic analysis of all the qualitative data from open-ended
questions collected across four years (2015--2018) from 21~teachers,
we identified eight distinct areas (indicated in bold) in which teachers
and schools seem to have been positively affected by their involvement
in PRiSE. These codes have subsequently been applied to responses
across all years of data (2015--2020, 45~teachers). Expressions
of negative impact were rare, with only one teacher (Teacher~15,
Tree Hill High School, MUSICS 2018) noting that the project had caused
them ``\textit{a small amount of extra stress}'', nonetheless, this
teacher continued to engage with the programme in subsequent years.
The first theme identified related to teachers \textbf{gaining new
physics knowledge} based around the content of the projects:\begin{quote}``{[}It{]}\textit{
added to }{[}my{]}\textit{ knowledge of standing waves giving more
real-life examples of waves.}'' (Teacher 8, Coal Hill School, MUSICS
2017)\\
 ``{[}It{]}\textit{ introduced me to an area of physics where I have
little experience. I have yet to teach the particles side of A-level
physics. However, this project and the knowledge accumulated will
be valuable when I do.}'' (Teacher~20, Hogwarts, SCREAM 2018)\end{quote}In
turn, this has made them\textbf{ more confident in discussing research}
with students in general:\begin{quote} ``{[}It has{]}\textit{ given
me confidence to explore physics beyond my areas of expertise / beyond
the school specs.}'' (Teacher~15, Spence Academy for Young Ladies,
MUSICS 2018)\\
``\textit{It has re-ignited my interest in current research, and
reminded me that complicated, cutting edge research can be more accessible
than I sometimes think!}'' (Teacher~36, Starfleet Academy, MUSICS
2020)\\
``{[}I have{]}\textit{ re-engaged with research and research methods.}''
(Teacher~41, Xavier's Institute for Higher Learning, SCREAM 2020)\end{quote}Elements
of the research projects have also been \textbf{implemented or referred
to in teachers' regular lessons}:\begin{quote}``{[}I have been{]}\textit{
able to use the detector with classes when teaching Year 12 particles.}''
(Teacher~13, St Trinians, SCREAM 2016)\textit{}\\
``\textit{It has been referred to }{[}in lessons{]}\textit{ in terms
of what scientists do and the research process.}'' (Teacher~15,
Spence Academy for Young Ladies, MUSICS 2018)\textit{}\\
\textit{``}{[}It has given me{]}\textit{ context when talking about
Earth's magnetic field }{[}in lessons{]}\textit{.}'' (Teacher~16,
Tree Hill High School, MUSICS 2018)\\
\textit{``It has consolidated my understanding and teaching of exoplanets.
I used some of the techniques in teaching detection of exoplanets
in the astro topic of AQA's A-level going beyond the syllabus.}''
(Teacher~19, Boston Bay College, PHwP 2018)\end{quote}Another theme
concerned teachers gaining \textbf{confidence in mentoring and/or
supporting extra curricular activities}, such as developing their
\textit{``patience }{[}and{]}\textit{ encouragement}'' (Teacher~6,
Hogwarts, SCREAM 2016):\begin{quote}``{[}I have developed in{]}\textit{
motivating students to attempt challenging problems.}'' (Teacher~11,
Prufrock Preparatory School , SCREAM 2017)\end{quote}They also report
developing a \textbf{variety of other skills} including algebra, data
analysis, reviewing academic posters, using software such as Audacity
and Excel, and computer programming:\begin{quote}``\textit{I have
also enjoyed the personal challenge to my own coding abilities.}''
(Teacher~39, Bending State College, PHwP 2020)\end{quote}\textbf{Students'
project work has been shared across the schools} such as via assemblies,
displaying students' posters in classrooms or halls within the school,
and publishing news stories on the school's website or in local papers
(references not included so as to preserve the anonymity of schools,
teachers, and students). We note though that this was communicated
to us informally either via email or during subsequent school visits
rather than through the paper questionnaire. Nonetheless it seemed
of sufficient note to include as a theme. Following on from this,
some teachers report in the survey that their school's involvement
with the project has \textbf{raised the profile of physics or STEM
within their school}:\begin{quote} ``{[}Students in lessons{]}\textit{
were impressed to hear of our `muon project' and knowing we were involved
with a university physics department helped them engage with us. If
they think they and their teachers can be involved in research they
are more motivated.}'' (Teacher~1, Hogwarts, SCREAM 2015)\\
``{[}It{]}\textit{ gave prestige to the Physics department at the
college.}'' (Teacher~9, Xavier's Institute for Higher Learning,
MUSICS 2017)\end{quote}Finally, some teachers feel that they and
their schools have \textbf{developed a relationship with the university}:\begin{quote}``{[}It{]}\textit{
created a link to HE.}'' (Teacher~15, Spence Academy for Young Ladies,
MUSICS 2018)\\
``\textit{We feel involved in a very interesting }{[}research{]}\textit{
project.}'' (Teacher~27, Hogwarts, SCREAM 2019)\end{quote}This
is further backed-up by the significant repeated buy-in of teachers
and schools after completing the project, with $70\pm10\%$ returning
for multiple years of PRiSE projects, further explored in \citet{archer_prise_schools20},
thereby changing how they interact with universities (e.g. not just
attending one-off events) and solidifying the above mentioned impacts:\begin{quote}
``\textit{I am more confident in my second year.}'' (Teacher~21,
Hogwarts, SCREAM 2018)\\
``\textit{Now I've done one project I feel better equipped to get
things going myself.}'' (Teacher~39, Bending State College, PHwP
2020)\end{quote}

We use these eight areas of impact on teachers and schools for subsequent
quantitative analysis in the next section. However, we note with further
teacher survey responses in 2019--2020 we have identified an additional
theme. This pertains to \textbf{teachers' preconceptions of their
students' ability}:\begin{quote}``\textit{I am now more aware of
what our students are capable of - not just listening to visiting
speakers but being actively engaged in real-world research!}'' (Teacher~10,
Prufrock Preparatory School, SCREAM 2017)\\
``{[}It has{]}\textit{ made me more enthusiastic to engage students
in real research.}'' (Teacher~17, Sunnydale High School, MUSICS
2018)\\
``\textit{The project allowed me to identify students that were genuinely
interested and committed to Physics. It also gave me evidence that
my students should study science further at university. I was able
to pass this on to parents and universities.}'' (Teacher~31, Quirm
College for Young Ladies {[}and partner schools{]}, MUSICS 2020)\\
``\textit{It has been inspiring to see my students self-organising
so well together.}'' (Teacher~43, Sunnydale High School, MUSICS
2020)\end{quote}

\subsection{Quantitative analysis}

Based on the areas of impact on teachers and schools emerging from
the qualitative data (from 2015--2018), from 2019 onwards we sought
to quantitatively assess how prevalent they might be. Teachers ($n=23$)
were asked to identify for each of the 8 themes whether they felt
that they (or their school) had been affected by the project in that
area, using the closed options of: ``I have'', ``I will eventually'',
``I have not'', and ``Unsure''. This scale was chosen over a 5-point
Likert due to an expected low level of responses. We exclude any blank
or unsure answers (which were rare) and divide the remaining 172 responses
across the 8 themes into negatives (``I have not'') and positives,
with the latter being subdivided into planned (``I will'') and definite
(``I have'') impacts. We acknowledge some may consider the ``I
will'' response as neutral and thus our analysis takes both interpretations
into account.

\begin{figure*}
\begin{centering}
\includegraphics{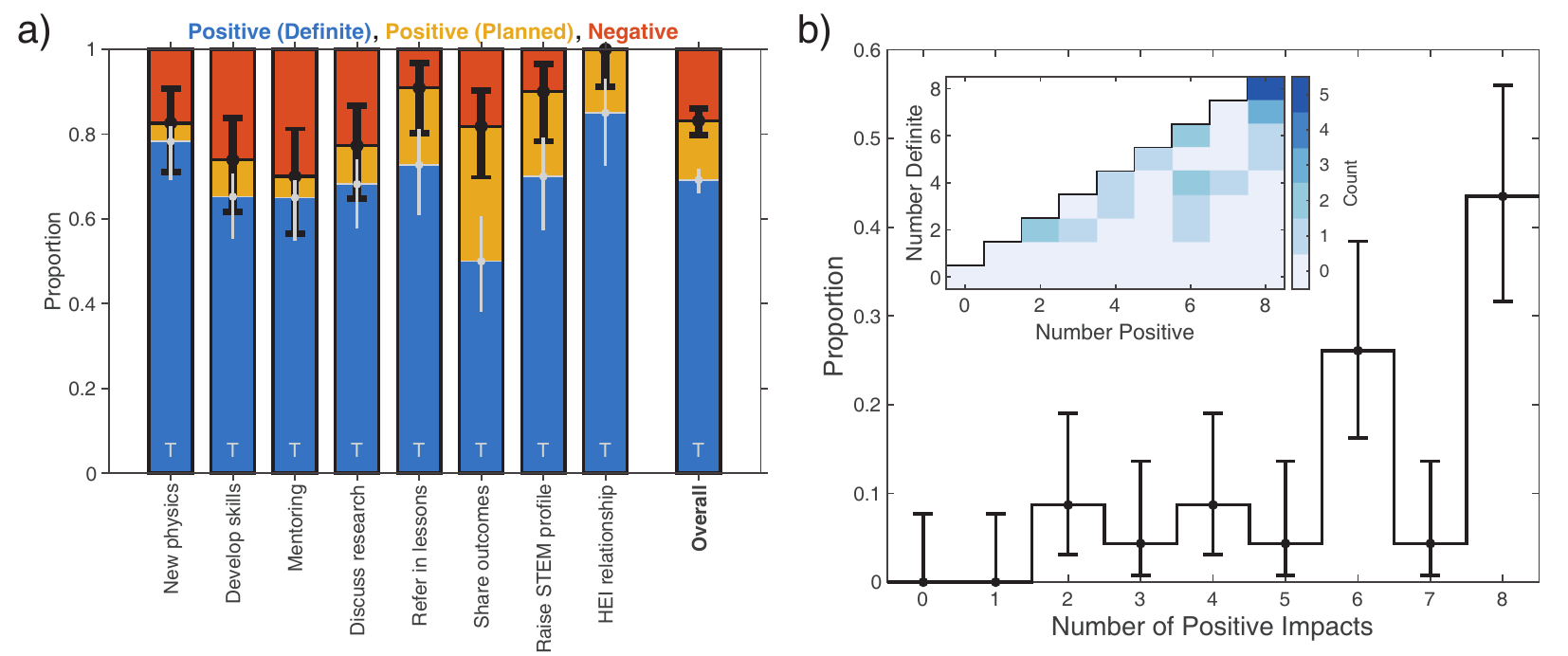}
\par\end{centering}
\caption{Quantitative results of impacts on teachers and schools ($n=23$).
a) Distribution of teachers' responses for each impact category. Results
are divided (black lines and associated error bars) into negative
(red) and positive responses, with the latter subdivided (grey lines
and error bars) into `definite' (blue) and `planned' (yellow). b)
Distribution of the number of positive impact categories reported
per teacher, with the bivariate distribution of the number of definite
impacts vs. positive ones shown inset. Error bars denote standard
(1$\sigma$) \citet{clopper34} intervals.\label{fig:teacher-impact}}
\end{figure*}

Figure~\ref{fig:teacher-impact}a shows the distributions of these
results for each impact area along with the overall results obtained
from totalling all responses. We find that all areas, apart from mentoring
($p=0.115$), have statistically significant positive majorities in
binomial tests ($p<0.03$). Only learning new physics and developing
a relationship with QMUL have majority definite responses ($p<0.011$).
Coding the responses to values of 1 (negative) to 3 (definite) the
overall average is $2.52\pm0.06$, which is greater than 2 to a very
high level of confidence. In fact all categories, apart from mentoring
($p=0.108$) and sharing outcomes ($p=0.119$), are statistically
significant in one-sample Wilcoxon signed-rank tests against null
hypotheses of 2. However, all categories' distributions are consistent
with the overall results in Wilcoxon rank sum tests, thus while there
are differences between the distributions such as slightly more teachers
feeling they have not developed mentoring skills or having not yet
shared their students' work across their schools (understandable given
the question was posed at the student conference) these variations
are moderate and not statistically significant.

Figure~\ref{fig:teacher-impact}b shows the distribution of the number
of positive impacts claimed per teacher. No teachers responded negatively
in all categories and thus all seem to have been positively affected
in some way. The average number of positive categories indicated by
teachers was $6.2\pm0.4$ out of 8. The bivariate distribution of
definite impacts per teacher as a function of the number of positive
impacts they indicated is shown inset, revealing teachers tended (apart
from in one case) to indicate the majority of positive impacts claimed
had already occurred. The average number of definite impacts was $5.2\pm0.5$.
Out of the 23 teachers, only 5 gave the same answer in all categories
(in these cases all definite responses) suggesting the results are
largely reliable and likely did not fall prey to unreflective responses.
Therefore, it appears that the identified areas of impact upon teachers
and schools as a result of PRiSE may indeed be quite widespread.

\conclusions{}

We have investigated the medium- and long-term impacts on students,
teachers, and schools who have participated in a 6-month-long programme
of physics `research in schools' projects, open-ended investigations
for school students based around cutting-edge STEM research. This
programme, `Physics Research in School Environments' (PRiSE), has
involved a diverse range of London schools and we have used evaluation
data captured from questionnaires across its entire 6~year duration
to date.

Medium-term impacts on the participating 14--18 year-old school students
were assessed after they had completed their 6-month-long projects.
Students' confidence in relevant scientific topics and methods seems
to have substantially increased as a result of PRiSE, with nearly
all students reporting this benefit. Furthermore, through experiencing
and being involved in research-level physics, students report having
gained new, or further developed existing, skills. Both of these impacts
upon students have been corroborated by teachers' observations. While
the students involved with PRiSE were fairly committed to STEM in
general beforehand, our data suggest that they had no clear bias in
aspirations towards the subject of physics in particular. Following
the programme it appears that students' attitudes towards pursuing
STEM were typically maintained or confirmed through their involvement
and physics aspirations seem to have been moderately enhanced. We
find no evidence that these impacts varied by the different projects
or schools. These results should be deemed successful, as a drop-off
in STEM aspirations is often seen at this age \citep{davenport20},
with these issues being particularly pertinent in physics \citep{archer20}.
Thus at this stage of a student's educational journey they are likely
to require interventions that sustain and support their science identity
which, in turn, has an influence on their educational choices \citep{archer16}.

Longitudinal evaluation has also been performed for three cohorts
of PRiSE students 3~years after they commenced their projects. While
a relatively small sample, the evidence suggests that these projects
have been memorable and beneficial experiences that students have
been able to draw upon in their later educational activities and development.
The data on PRiSE students' degree destinations show increased uptake
of physics and STEM at degree-level than would typically be expected,
suggesting that their involvement in the research projects has helped
transform their aspirations into destinations --- a key aim of the
programme. Further in-depth qualitative research, such as interviews
or focus groups, could provide richer and more reflective information
on how students' thoughts and feelings about their association with
physics and STEM may have been affected by participating in PRiSE,
given the nuance and multiple factors at play with students' aspirations
in general, which are difficult to capture and interpret with questionnaire
data alone \citep{aspires13,aspires20}.

The impacts upon students reported in this paper only relate to those
who completed the 6-month programme. However, as should be expected
for any extended programme, there is some drop-off in participation
with PRiSE. This has been explored in more detail in \citet{archer_prise_schools20},
demonstrating that students' success with PRiSE appears to be independent
of background and thus not clearly patterned by societal biases present
within the field \citep[e.g.][]{case14}. Currently though we have
no data on what impact the programme has on those students who drop
out. While no teachers have communicated any negative effects on students
who do not continue, with some highlighting informally their students'
attitudes towards the projects, this requires further formal investigation.
Such work is required to ensure that no negative effects are being
felt by these students and potentially discover what positives, if
any, may result from even partial participation. Furthermore, we have
no evidence that the PRiSE approach would be effective for students
who are generally uninterested or unengaged with STEM. Indeed, it
seems unlikely that such students would want to persist with an extended
and challenging extra-curricular physics programme. Young people's
aspirations towards science begin to form at an early age \citep{aspires13,aspires20}
and therefore interventions throughout their educational journey need
to align with their needs and wants, from initial inspiration and
positive associations, to informing on career-focused aspects, and
finally sustaining those built aspirations \citep{davenport20}. PRiSE
only aims to address that final part of the chain, since no single
programme can fit all stages.

Evaluation of the impacts on teachers and schools has identified several
themes. By collaborating on PRiSE, teachers can gain new physics knowledge,
become more confident in discussing research, and integrate aspects
of the research projects into their regular lessons. Teachers also
report developing various technical skills, gaining confidence in
mentoring, and reassessing their preconceptions of students' potential.
While all these positive changes to teachers' practice will likely
be felt across their wider schools, there is more direct evidence
of the school environment being affected such as through students'
project work being championed, the profile of physics or science being
raised, and a university--school relationship being established with
significant repeated buy-in from schools over several years. These
impacts appear to be fairly widespread across the teachers and schools
involved in PRiSE. We note that these results share many similarities
to those reported by \citet{rushton19} for \citet{iris20} from interviews
with 17 teachers. The PRiSE programme features much greater diversity
in schools, with an over-representation of disadvantaged groups in
many metrics considered \citep{archer_prise_schools20}. Furthermore,
as discussed in \citet{archer_prise_framework20}, PRiSE does not
rely on such a teacher-driven model instead providing a wealth of
resources and interventions to support teachers' and schools' participation.
The similar impacts thus highlight that, with the right support, teachers
and schools from a variety of contexts can benefit from `research
in schools' projects. Further research could investigate the validity
of the participating teachers' remarks of the impact on the schools'
environments, for instance through in-depth interviews or focus groups
with other teachers in the schools.

The impacts upon participating students, teachers, and schools discussed
in this paper show real promise for the emerging field of `research
in schools' initiatives. They suggest that with more similarly designed
and supported programmes at other institutions, we may be able to
start to address a key part of the chain of the wider issue of uptake
and diversity not just in physics but potentially STEM also. We stress,
however, that multi-faceted approaches from a variety of different
stakeholders and organisations are required to implement real change
on this entire issue, but `research in schools' may be able to form
one piece of the puzzle.

\appendix

\section{6-month stage evaluation questions\label{sec:6-month-questions}}

Here we list the questions considered within this paper posed in the
PRiSE-wide questionnaires at the 6-month stage evaluation. We detail
the phrasing used, how participants could respond, and which years
the question was asked. Follow-on questions are indicated by indentation
and a down-right arrow (\rotatebox[origin=c]{180}{$\Lsh$}). The following
questions were posed to students:

\begin{footnotesize}%
\begin{longtable}[c]{>{\raggedright}p{0.5\paperwidth}cc}
\textbf{Question} & \textbf{Response type} & \textbf{Year(s)}\tabularnewline
\hline 
\noalign{\vskip8pt}
In what way has this project affected you & Open Text & 2016--2020\tabularnewline
\noalign{\vskip8pt}
What skills, if any, has the project helped you develop & Open Keywords & 2016--2020\tabularnewline
\noalign{\vskip8pt}
How has doing the project affected your thoughts about future subject
choices / careers & Open Text & 2018--2019\tabularnewline
\noalign{\vskip8pt}
Before working on the project, how likely were you to continue with
the following (Physics/STEM) in the future & 5-point Likert & 2020\tabularnewline
\noalign{\vskip8pt}
\setlength{\leftskip}{1em}\rotatebox[origin=c]{180}{$\Lsh$}After
working on the project, how likely are you now to continue with the
following (Physics/STEM) in the future & 5-point Likert & 2020\tabularnewline
\noalign{\vskip8pt}
\setlength{\leftskip}{1em}\rotatebox[origin=c]{180}{$\Lsh$}How has
working on the project affected your thoughts about these future subject
choices (Physics/STEM) & 5-point Likert & 2020\tabularnewline
\noalign{\vskip8pt}
\setlength{\leftskip}{2em}\rotatebox[origin=c]{180}{$\Lsh$}Please
explain how or why? & Open Text & 2020\tabularnewline
\noalign{\vskip8pt}
\end{longtable}\end{footnotesize}The questions asked of teachers were:\begin{footnotesize}%
\begin{longtable}[c]{>{\raggedright}p{0.5\paperwidth}cc}
\textbf{Question} & \textbf{Response type} & \textbf{Year(s)}\tabularnewline
\hline 
\noalign{\vskip8pt}
In what way has this project affected your students & Open Text & 2015--2020\tabularnewline
\noalign{\vskip8pt}
In what way has this project affected you & Open Text & 2015--2018\tabularnewline
\noalign{\vskip8pt}
What skills, if any, has the project helped your students develop & Open Keywords & 2016--2020\tabularnewline
\noalign{\vskip8pt}
What skills, if any, has the project helped you develop & Open Text & 2016--2018\tabularnewline
\noalign{\vskip8pt}
Have you found the project useful in your lessons & 5-point Likert & 2015--2018\tabularnewline
\noalign{\vskip8pt}
\setlength{\leftskip}{1em}\rotatebox[origin=c]{180}{$\Lsh$}Please
tell us why / why not & Open Text & 2015--2018\tabularnewline
\noalign{\vskip8pt}
Do you feel you have been affected by the project in any of the following
ways (8 categories) & Closed Options & 2019--2020\tabularnewline
\noalign{\vskip8pt}
\setlength{\leftskip}{1em}\rotatebox[origin=c]{180}{$\Lsh$}Are there
any other ways you feel you've been affected & Open Text & 2019--2020\tabularnewline
\noalign{\vskip8pt}
\end{longtable}\end{footnotesize}

\section{3-year stage evaluation questions\label{sec:3-year-questions}}

The following questions were asked of students in the 3-year stage
evaluation via an online form.

\begin{footnotesize}%
\begin{longtable}[c]{l>{\raggedright}p{0.3\paperwidth}cc>{\centering}p{0.08\paperwidth}c}
\textbf{Section} & \textbf{Question} & \textbf{Response type} & \textbf{Required} & \textbf{Go to section} & \textbf{Cohort(s)}\tabularnewline
\hline 
\noalign{\vskip8pt}
1 & What school year would you have been in during the academic year {[}cohort
year{]} & Closed Options & Y &  & 1--3\tabularnewline
\noalign{\vskip8pt}
 & Do you remember undertaking a Queen Mary Physics \& Astronomy research
project that year & Yes/No & Y & Yes: 2\\
No: 3 & 1--3\tabularnewline
\hline 
\noalign{\vskip8pt}
2 & What experiences do you remember from the project & Open Text & N &  & 1--3\tabularnewline
\noalign{\vskip8pt}
 & How have you used those experiences since, if at all & Open Text & N & 3 & 1--3\tabularnewline
\hline 
\noalign{\vskip8pt}
3 & Are you studying at a Higher Education Institution (HEI) e.g. a University & Closed Options & Y & Yes: 4\\
Intend to: 5\\
No: Submit & 1--3\tabularnewline
\hline 
\noalign{\vskip8pt}
4 & Which HEI are you studying at & Open Text & N & 5 & 1--3\tabularnewline
\hline 
\noalign{\vskip8pt}
5 & What subject are you studying or plan to study at a HEI & Closed Options & Y &  & 1--3\tabularnewline
\noalign{\vskip8pt}
 & What are/were your reasons for studying that subject & Open Text & N &  & 1--3\tabularnewline
\noalign{\vskip8pt}
 & What influenced you to study that subject & Open Text & N & 6 & 1--3\tabularnewline
\hline 
\noalign{\vskip8pt}
6 & Did the research project in any way influence your subject choice & Closed Options & Y &  & 2--3\tabularnewline
\noalign{\vskip8pt}
 & How would you say the research project affected your subject choice
(if applicable) & Open Text & N & Submit & 2--3\tabularnewline
\hline 
\noalign{\vskip8pt}
\end{longtable}

\end{footnotesize}

\dataavailability{Data supporting the findings of this study that
is not already contained within the article or derived from listed
public domain resources are available on request from the corresponding
author. This data is not publicly available due to ethical restrictions
based on the nature of this work.}

\authorcontribution{MOA conceived the programme and its evaluation,
performed the analysis, and wrote the paper. JDW contributed towards
the analysis, validation, and writing.}

\competinginterests{The authors declare that they have no conflict
of interest.}
\begin{acknowledgements}
We thank Dominic Galliano, Olivia Keenan, and Charlotte Thorley for
helpful discussions. MOA holds a UKRI (STFC / EPSRC) Stephen Hawking
Fellowship EP/T01735X/1 and received funding from the Ogden Trust.
This programme has been supported by a QMUL Centre for Public Engagement
Large Award, and STFC Public Engagement Small Award ST/N005457/1.
\end{acknowledgements}
\DeclareRobustCommand{\disambiguate}[3]{#1}\bibliographystyle{copernicus}
\bibliography{prise2020}

\end{document}